\documentclass[aps,prd,preprint,preprintnumbers,nofootinbib,superscriptaddress]{revtex4}
\usepackage{ae}
\usepackage{bm}
\usepackage{amssymb}
\usepackage{amsmath}
\usepackage{amsfonts}
\usepackage{graphicx}
\usepackage{slashed}
\usepackage{wasysym}
\usepackage{setspace}
\usepackage{ulem}
\usepackage{multirow}
\usepackage{footnote}
\usepackage{upgreek}
\usepackage{siunitx}

\newcommand{\Eqref}[1]{Eq.~\eqref{#1}}

\allowdisplaybreaks

\begin{document}

\setlength{\unitlength}{1mm}

\title{Vacuum birefringence at x-ray free-electron lasers} 


\author{Felix Karbstein}\email{felix.karbstein@uni-jena.de}
\affiliation{Helmholtz-Institut Jena, Fr\"obelstieg 3, 07743 Jena, Germany}
\affiliation{Theoretisch-Physikalisches Institut, Abbe Center of Photonics, \\ Friedrich-Schiller-Universit\"at Jena, Max-Wien-Platz 1, 07743 Jena, Germany}
\affiliation{GSI Helmholtzzentrum f\"ur Schwerionenforschung, Planckstr. 1, 64291 Darmstadt, Germany}
\author{Chantal Sundqvist}
\affiliation{Helmholtz-Institut Jena, Fr\"obelstieg 3, 07743 Jena, Germany}
\affiliation{Theoretisch-Physikalisches Institut, Abbe Center of Photonics, \\ Friedrich-Schiller-Universit\"at Jena, Max-Wien-Platz 1, 07743 Jena, Germany}	
\author{Kai S. Schulze}
\affiliation{Helmholtz-Institut Jena, Fr\"obelstieg 3, 07743 Jena, Germany}
\affiliation{Institut f\"ur Optik und Quantenelektronik, Friedrich-Schiller-Universit\"at Jena, Max-Wien-Platz 1, 07743 Jena, Germany}
\affiliation{GSI Helmholtzzentrum f\"ur Schwerionenforschung, Planckstr. 1, 64291 Darmstadt, Germany}
\author{Ingo Uschmann}
\affiliation{Helmholtz-Institut Jena, Fr\"obelstieg 3, 07743 Jena, Germany}
\affiliation{Institut f\"ur Optik und Quantenelektronik, Friedrich-Schiller-Universit\"at Jena, Max-Wien-Platz 1, 07743 Jena, Germany}
\author{Holger Gies}
\affiliation{Helmholtz-Institut Jena, Fr\"obelstieg 3, 07743 Jena, Germany}
\affiliation{Theoretisch-Physikalisches Institut, Abbe Center of Photonics, \\ Friedrich-Schiller-Universit\"at Jena, Max-Wien-Platz 1, 07743 Jena, Germany}
\author{Gerhard G. Paulus}
\affiliation{Helmholtz-Institut Jena, Fr\"obelstieg 3, 07743 Jena, Germany}
\affiliation{Institut f\"ur Optik und Quantenelektronik, Friedrich-Schiller-Universit\"at Jena, Max-Wien-Platz 1, 07743 Jena, Germany}

\date{\today}

\begin{abstract}
 We study the perspectives of measuring the phenomenon of vacuum birefringence predicted by quantum electrodynamics using an x-ray free-electron laser (XFEL) alone.
 We devise an experimental scheme allowing two consecutive XFEL pulses to collide under a finite angle, and thus act as both pump and probe field for the effect.
 The signature of vacuum birefringence is encoded in polarization-flipped signal photons to be detected with high-purity x-ray polarimetry.  
 Our findings for idealized scenarios underline that the discovery potential of solely XFEL-based setups can be comparable to those involving optical high-intensity lasers.
 For currently achievable scenarios, we identify several key details of the x-ray optical ingredients that exert a strong influence on the magnitude of the desired signatures. 
\end{abstract}

\maketitle

\section{Introduction}

Vacuum birefringence is a finger print of quantum vacuum nonlinearity in strong electromagnetic fields \cite{Toll:1952,Baier,BialynickaBirula:1970vy,Adler:1971wn}:
when a strong electromagnetic field featuring a preferred direction is applied to the quantum vacuum, it can effectively supplement the latter with two different indices of refraction. 
Probe light with overlap to both of these polarization modes can experience a birefringence phenomenon. Originally linearly polarized probe light becomes slightly elliptically polarized, and thus gives rise to polarization-flipped signal photons.

The microscopic origin of this effect is the omnipresence of quantum vacuum fluctuations. Within the Standard model of particle physics, the dominant contribution arises from the effective self-coupling of electromagnetic fields mediated by an electron-positron fluctuation \cite{Euler:1935zz,Heisenberg:1935qt} as described by quantum electrodynamics (QED).
The latter supplements Maxwell's linear theory of electrodynamics with non-linear corrections. These are, however, very small for the electric $E$ and magnetic $B$ fields available in the laboratory, generically fulfilling $\{E,cB\}\ll E_{\rm crit}$, with {\it critical} electric field $E_{\rm crit}=\frac{m_e^2c^3}{e\hbar}\simeq1.3\times10^{18}\,\frac{\rm V}{\rm m}$.
Hence, though having been predicted already in the early days of quantum electrodynamics and being actively searched for already since several decades, quantum vacuum nonlinearities in macroscopic electromagnetic fields have so far not been observed in a controlled laboratory experiment. 

Ongoing laboratory searches for vacuum birefringence, such as the PVLAS \cite{Ejlli:2020yhk}, BMV \cite{Cadene:2013bva,Hartman:2017nez} and OVAL \cite{Fan:2017fnd} experiments, use continuous wave lasers as a probe and quasi-static magnetic fields to polarize the quantum vacuum in combination with high-finesse cavities to increase the effective interaction length; cf. \cite{Battesti:2018bgc} for a recent review.
Especially in the last decades, the progress in the development of petawatt-class lasers and brilliant XFEL light sources have triggered various theoretical proposals \cite{Aleksandrov:1985,Kotkin:1996nf,Heinzl:2006xc,DiPiazza:2006pr,Dinu:2013gaa,Dinu:2014tsa,Nakamiya:2015pde,Karbstein:2015xra,Ilderton:2016khs,King:2016jnl,Schlenvoigt:2016,Karbstein:2016lby,Bragin:2017yau,Karbstein:2018omb,Ataman:2018ucl,Mosman:2021vua} to measure the effect in an all-optical experiment, where both the probe and the pump fields are generated by lasers; cf. \cite{DiPiazza:2011tq,King:2015tba,Karbstein:2019oej} for recent reviews.
A particularly prominent scenario is the head-on collision of a high-intensity laser pulse with a bright XFEL pulse. This requires the co-location of an XFEL and a petawatt-class high-intensity laser. In fact, such experiments are planned at the Helmholtz International Beamline for Extreme Fields (HIBEF) \cite{HIBEF} at the European XFEL \cite{XFEL}, as well as SACLA \cite{Inada:2017lop,Seino:2019wkb} and the Shanghai Coherent Light Facility \cite{Shen:2018lbq}.

In the present article, we study the perspectives of detecting vacuum birefringence solely based on the availability of an XFEL light source.
The underlying idea is to introduce a self-crossing of the XFEL beam, allowing for the collision of two consecutive pulses.
More precisely, we envision to make use of the well-controlled, regular pulse structure of the XFEL beam and adjust the propagation distances in our setup such as to ensure the collision of subsequent XFEL pulses.
Hence, a given pulse acts as both pump and probe. After serving as a pump pulse for the preceding pulse, it probes the subsequent one.

Our work is motivated by two facts: first, the observation that the peak field strength reachable with a tightly focused XFEL pulse seems to approach those attainable with high-intensity lasers, and, second, the high-repetition rates of upcoming XFELs based on superconducting cavities. 
In order to illustrate the first point, we compare the peak fields reached by the high-intensity laser installed at the HIBEF beamline at the European XFEL with that achievable by a tightly focused XFEL beam.
To this end, we assume both laser fields to be well-described as pulsed fundamental Gaussian beams.
The HIBEF laser delivers pulses of energy $W=10\,{\rm J}$ at a wavelength of $\lambda=800\,{\rm nm}$ and a pulse duration of $\tau=25\,{\rm fs}$, which can be focused to a waist spot size of $w_0\simeq1\,\si\micro{\rm m}$ at a repetition-rate of $5\,{\rm Hz}$.
For an XFEL with photon energy $\omega_{\rm XFEL}$, pulse duration $\tau_{\rm XFEL}$ and number of photons per pulse $N_{\rm XFEL}$, the same peak intensity is reached if the XFEL is focused to a waist of 
$w_{0,{\rm XFEL}}=w_0\sqrt{\frac{\tau}{\tau_{\rm XFEL}}\frac{N_{\rm XFEL}\omega_{\rm XFEL}}{W}}$. For the parameters of the HIBEF laser and resorting to the commonly used estimates $N_{\rm XFEL}\simeq10^{12}$, $\omega_{\rm XFEL}\simeq9\,{\rm keV}$ and $\tau_{\rm XFEL}=\tau$, this expression results in $w_{0,{\rm XFEL}}\approx12\,{\rm nm}$. In fact, XFEL beam waists of the order of $100\,{\rm nm}$ have already been demonstrated experimentally \cite{Mimura:2014}, and waists of the order of $10\,{\rm nm}$ should be feasible \cite{Yamauchi:2011,Huang:2013}.
Concerning the second point, we just note that the repetition rate of the European XFEL is as large as $27\,000\,{\rm pulses}/{\rm s}$ \cite{XFEL}.
Hence, even for a smaller signal photon yield per shot in comparison to an XFEL plus high-intensity laser scenario, the actual number of events per given time interval in a purely XFEL based setup may be of the same order or even larger than the former.

Our article is structured as follows. After briefly recalling the theoretical foundations, we detail our calculation and the underlying assumptions in Sec.~\ref{sec:calc}.
Here, one of our central goals is to obtain a tractable analytical expression for the angularly resolved differential number of polarization-flipped signal photons, which can be readily evaluated for different parameter regimes.
In Section~\ref{sec:setup} we propose and describe an experimentally feasible setup allowing to perform such an experiment at an XFEL source.
First, in Sec.~\ref{subsec:Idealized}, we evaluate the attainable numbers of polarization-flipped signal photons under idealized assumptions typically invoked in the literature, neglecting in particular the losses and pulse deformations induced by the optic elements inserted into the beam path. The results of this analysis look rather promising and suggest the possibility of a measurement of the effect in an experiment with state-of-the-art technology.
Second, in Sec.~\ref{subsec:Realistic}, we estimate the relevant losses, repeat the analysis, and find a drastic reduction of the signal for present-day parameters.
A comparison of these results underpins how predictions based on seemingly realistic assumptions can significantly be altered when accounting for the details of the experimental setup devised to measure the effect. In this way, we identify a number of key issues where an optimization of technical ingredients can lead to substantial improvement of the experimental discovery potential.
Finally, in Sec.~\ref{sec:outlook} we {provide a brief outlook on potential ways to enhance the signal in experiment and end with} conclusions in Sec.~\ref{sec:concls}.

\section{Calculation}
\label{sec:calc}

We analyze the attainable number of polarization-flipped signal photons encoding the signature of vacuum birefringence using the vacuum emission picture \cite{Galtsov:1971xm,Karbstein:2014fva}, recasting all-optical signatures of quantum vacuum nonlinearity in prescribed macroscopic electromagnetic fields as signal photon emission processes; see Ref.~\cite{Karbstein:2019oej} for a recent review. Throughout this work, we use units where $c=\hbar=1$.

Vacuum birefringence amounts to a single signal photon emission process.
It is typically analyzed in terms of a pump-probe type scenario: the former effectively polarizes the quantum vacuum and the latter probes this excited space-time region. The experimental signature of the effect are polarization-flipped signal photons originating from the probe beam. 
This implies that the associated signal photon current is linear in the probe photon field.
Because the leading QED vacuum nonlinearity is an effective four field interaction, it is moreover quadratic in the pump field.  

As the self-crossing of an XFEL beam generically happens under a finite angle (cf. Fig.~\ref{fig:setup} below), we have to study the effect accounting for a finite collision angle of $\vartheta_{\rm coll}$.
In the present derivation, we keep the calculation a bit more general and consider the collision of two independent laser beams. These results can then readily be applied to the special case which is relevant here. 

\subsection{Collision geometry}

Without loss of generality, we assume the beam axes of the two driving lasers to lie in the xz-plane and their unit wave vectors to be given by $\hat{\vec{\kappa}}_1=(0,0,1)$ and $\hat{\vec{\kappa}}_2=(\sin\vartheta_{\rm coll},0,\cos\vartheta_{\rm coll})$.
The possible linear polarizations of each of these beams $b\in\{1,2\}$ can then be characterized by a single angle $\beta_b$, and the associated polarization vectors read $\hat{\vec{E}}_{1}=(\cos\beta_1,\sin\beta_1,0)$ and $\hat{\vec{E}}_{2}=(\cos\vartheta_{\rm coll}\cos\beta_2,\sin\beta_2,-\sin\vartheta_{\rm coll}\cos\beta_2)$, respectively.
On the other hand, the wave vectors of the signal photons can be parameterized by $\vec{k}={\rm k}\hat{\vec{k}}$, with $\hat{\vec{k}}=(\cos\varphi\sin\vartheta,\sin\varphi\sin\vartheta,\cos\vartheta)$, and the associated transverse polarization vectors by $\hat{\vec{e}}_{\beta}=\sin\beta\,\hat{\vec{k}}|_{\vartheta=\frac{\pi}{2},\varphi\to\varphi+\frac{\pi}{2}}+\cos\beta\,\hat{\vec{k}}|_{\vartheta\to\vartheta+\frac{\pi}{2}}$.

The corresponding results for the signal photon transition amplitude ${\cal S}_\beta(\vec{k})$ to a state where the polarization vector of the signal photon of momentum $\vec{k}$ is given by $\hat{\vec{e}}_{\beta}$ follow straightforwardly from Eqs.~(19)-(21) of Ref.~\cite{Gies:2017ygp} upon identifying $\vartheta_2\to\vartheta_{\rm coll}$.
For the vacuum birefringence study performed here, we assume beam $b=1$ to constitute the probe and beam $b=2$ the pump, and thus limit ourselves to the contribution to ${\cal S}_\beta(\vec{k})$ which is linear in the probe field. This results in
\begin{align}
{\cal S}_{\beta}(\vec{k})=&{\rm i} \frac{\sqrt{\alpha}}{(2\pi)^{3/2}}\frac{1}{45} 
\,m_e^2\sqrt{\rm k}\,(1-\cos\vartheta_{\rm coll})  \nonumber\\
&\times\Bigl\{ 
\bigl[\cos\varphi
(1-\cos\vartheta\cos\vartheta_{\rm coll})-\sin\vartheta\sin\vartheta_{\rm coll}
\bigr]  f(\beta_1+\beta_2,\beta+\beta_2) \nonumber\\
&\quad\quad-\sin\varphi(\cos\vartheta-\cos\vartheta_{\rm coll})\, 
f(\beta_1+\beta_2,\beta+\beta_2-\tfrac{\pi}{2})\Bigr\}\,{\cal I}_{12}(k) \, , \label{eq:Sfinal}
\end{align}
with fine structure constant $\alpha=e^2/4\pi\simeq1/137$.
Here, we made use of the shorthand notations
\begin{equation}
 f(\mu,\nu)=4\cos\mu\cos\nu+7\sin\mu\sin\nu \,,  \label{eq:shorthands}
\end{equation}
and
\begin{equation}
 {\cal I}_{12}(k)=\int{\rm d}^4 x\,{\rm e}^{{\rm i} {\rm k}(\hat{\vec 
k}\cdot\vec{x}-t)}\,\frac{e{\cal E}_1(x)}{m_e^2}\Bigl(\frac{e{\cal E}_2(x)}{m_e^2}\Bigr)^2\,. \label{eq:I12}
\end{equation}
The functions introduced in \Eqref{eq:shorthands} encode the dependence of \Eqref{eq:Sfinal} on the probe, pump and signal polarizations, and \Eqref{eq:I12} its dependence on the field profiles of the driving laser beams measured in units of the QED critical field; recall that $E_{\rm cr}=m_e^2/e$. 

As noted above, the signature of vacuum birefringence amounts to probe photons scattered into a perpendicularly polarized mode. 
To isolate this signal, we choose the angle $\beta$ characterizing the polarization of the signal photons such that the condition $\hat{\vec{e}}_\beta\cdot\hat{\vec{E}}_{1}=0$ is fulfilled.
This results in $\beta\to\beta_\perp=\arctan\{\cos\vartheta\cot(\varphi-\beta_1)\}$. We denote the associated signal photon transition amplitude by ${\cal S}_\perp(\vec{k})={\cal S}_{\beta}(\vec{k})|_{\beta\to\beta_\perp}$. From this result, the differential number of polarization-flipped signal photons to be detected far outside the interaction region of the driving laser fields follows as
\begin{equation}
 {\rm d}^3N_\perp=\frac{{\rm d}^3k}{(2\pi)^3}\bigl|{\cal S}_\perp(\vec{k})\bigr|^2\,.
 \label{eq:dNperp}
\end{equation}

Subsequently, we model the driving laser fields as pulsed paraxial Gaussian beams and limit ourselves to the special case of two beams of the same oscillation frequency $\omega$.
Being interested in the principle possibility of such an experiment with present technology, throughout this work, we only consider ideal collisions characterized by an optimal temporal synchronization and perfectly overlapping foci of the two beams.
The incorporation of non-optimal conditions is important for modeling an actual experiment and can, e.g., be performed along the lines of Ref.~\cite{Schlenvoigt:2016}.

At the same time, the far-field angular decay of the number of the driving laser photons $N_b\simeq\frac{W_b}{\omega}$ with the polar angle $\vartheta_b$ measured from the beam's forward beam axis should be well-described by the Gaussian beam behavior \cite{Karbstein:2019oej}
\begin{equation}
 \frac{{\rm d}N_b}{{\rm d}\!\cos\vartheta_b}\simeq N_b(\omega w_{0,b})^2\,{\rm e}^{-\frac{1}{2}(\omega w_{0,b})^2\vartheta_b^2} \,. \label{eq:decaydriver}
\end{equation}
For the scenario discussed here, we are interested in the comparison of the angular decay of the vacuum birefringence signal and the probe photons of beam $b=1$ traversing the interaction region unaltered \cite{Karbstein:2015xra}.
Our coordinates are such that $\vartheta_1=\vartheta$.

\subsection{Infinite Rayleigh range approximation}
\label{sec:infiniteRayleigh}

To simplify the required computational efforts in the evaluation of ${\cal S}_\perp(\vec{k})$, we formally send the Rayleigh ranges ${\rm z}_{R,b}$ of the driving beams to infinity when determining the signal photon emission amplitude \cite{Gies:2017ygp,Karbstein:2019oej}.
In this limit, the field profiles of the driving laser beams $b\in\{1,2\}$ simplify considerably and read
\begin{equation}
 {\cal E}_b(x)={\mathfrak E}_b\,{\rm e}^{-\bigl(\frac{\vec{x}\cdot\hat{\vec{\kappa}}_{b}-t}{\tau_b/2}\bigr)^2} {\rm e}^{-\frac{\vec{x}^2-(\vec{x}\cdot\hat{\vec{\kappa}}_{b})^2}{w_{0,b}^2}} \cos\{\omega(\vec{x}\cdot\hat{\vec{\kappa}}_{b}-t)\}\,. \label{eq:E}
\end{equation}
The peak field amplitude ${\mathfrak E}_b$ can be expressed in terms of the laser pulse energy $W_b$, pulse duration $\tau_b$ and waist size $w_{0,b}$ via \cite{Karbstein:2017jgh}
\begin{equation}
 {\mathfrak E}_{b}\simeq 2\Bigl(\frac{8}{\pi}\Bigr)^{\frac{1}{4}}\sqrt{\frac{W_b}{\tau_b w_{0,b}^2\pi}}\,. 
\end{equation}
Here, $\tau_b$  and $w_{0,b}$ denote the pulse duration and beam waist measured at $1/{\rm e}^2$ of its peak intensity, respectively.
In an experimental context, these quantities are typically specified in terms of (half-)widths at half maximum (HM) on the level of the intensity.
These are related to the $1/{\rm e}^2$ parameters as
\begin{equation}
 \tau_{b}=\tau_b^{\rm HM}\sqrt{\frac{2}{\ln 2}} \quad\quad \text{and} \quad\quad w_{0,b}=w_{0,b}^{\rm HM}\sqrt{\frac{2}{\ln 2}}\,,
\end{equation}
where $\sqrt{2/\ln 2}\simeq1.7$.

The signal photons predominately originate from the strong-field region where the driving laser beams overlap, constituting the interaction region.
This implies that only local information about the driving fields is needed for determining the signal.
Correspondingly, the infinite Rayleigh range approximation is well-justified as long as the field profiles in the interaction region are essentially insensitive to variations of the Rayleigh ranges ${\rm z}_{R,b}$ of the driving beams $b\in\{1,2\}$.
This requires the existence of an additional physical length scale which limits the spatial extent of the interaction region along the beam axes of the driving laser fields to scales much smaller than ${\rm z}_{R,b}$.

For exactly counter-propagating laser fields, the only candidate for such a scale are the pulse durations $\tau_b$, implying the infinite Rayleigh range approximation to allow for reliable insights given that $\tau_b\ll{\rm z}_{R,b}$; cf. also Ref.~\cite{King:2018wtn}.
This is different for collisions under a nontrivial angle $0^\circ<\vartheta_{\rm coll}<180^\circ$, because $\vartheta_{\rm coll}$ can induce additional finite length scales along the propagation directions of the beams.  
From elementary geometric considerations it is clear that for the case of two beam collisions as considered here, this gives rise to the additional criterion 
\begin{equation}
 \Bigl\{\frac{w_{0,1}}{{\rm z}_{R,2}},\frac{w_{0,2}}{{\rm z}_{R,1}}\Bigr\}\ll|\sin\vartheta_{\rm coll}|\,.
 \label{eq:criterion} 
\end{equation}
Hence, if at least one of these criteria is met, a calculation resorting to the infinite Rayleigh range approximation should reproduce the results of an exact calculation employing self-consistent paraxial Gaussian beams with high accuracy. 

While the criterion in \Eqref{eq:criterion} formally involves a much-less-than ``$\ll$" sign, the findings in Ref.~\cite{Gies:2017ygp} suggest the infinite Rayleigh range approximation to be reliable up to the limit where the quantity on the left hand side of \Eqref{eq:criterion} reaches the size of its right hand side.
In fact, in the full parameter regime $\{\frac{w_{0,1}}{{\rm z}_{R,2}},\frac{w_{0,2}}{{\rm z}_{R,1}}\}\lesssim|\sin\vartheta_{\rm coll}|$, the deviations from a calculation using self-consistent paraxial Gaussian beams should be below the $10\%$ level.
This accuracy is sufficient for the present exploratory study.

For x-ray beams of photon energy $\omega=9835\,{\rm eV}$ and beam waists of the order of $w_0\simeq100\,{\rm nm}$ ($10\,{\rm nm}$) as considered here, we have ${\rm z}_{R}=\frac{w_{0}^2\omega}{2}$, such that $\frac{w_0}{{\rm z}_R}\simeq4\times10^{-4}$ ($4\times10^{-3}$).
Hence, the condition $\{\frac{w_{0,1}}{{\rm z}_{R,2}},\frac{w_{0,2}}{{\rm z}_{R,1}}\}\lesssim|\sin\vartheta_{\rm coll}|$ results in a restriction of the possible ranges of collision angles which can be reliably studied based on the infinite Rayleigh range approximation. 
For the parameters considered here, these collision angles are constrained by $0.02^\circ\leq\vartheta_{\rm coll}\leq179.98^\circ$ ($0.2^\circ\leq\vartheta_{\rm coll}\leq179.8^\circ$). These angular regimes cover all relevant collision angles:
a smaller collision angle near $\vartheta_{\rm coll}\simeq0$ is not of interest at all as the effect decreases rapidly towards $\vartheta_{\rm coll}\to0$, and eventually vanishes for co-propagating beams.
Conversely, a larger collision angle near $\vartheta_{\rm coll}\simeq180^\circ$ would maximize the signal, but presently does not seem to be experimentally feasible based on the availability of only a single free-electron laser at a given site; cf. Sec.~\ref{sec:setup} below.

Given that this criterion is fulfilled for all collision angles which are of interest to us, we do not even need to bother about the alternative criterion $\tau_b\ll{\rm z}_{R,b}$.
Note, however, that we do have $\tau\lesssim{\rm z}_R$ for pulse durations fulfilling $\tau\lesssim834\,{\rm fs}$ ($8.3\,{\rm fs}$).

\subsection{Signal photon numbers}

We perform the Fourier integrations in \Eqref{eq:I12} upon insertion of \Eqref{eq:E} and limit ourselves to the dominant quasielastic contribution. This gives rise to signal photons of energy $\approx\omega$ scattered in the vicinity of the positive $\rm z$ direction \cite{Karbstein:2015xra,Karbstein:2016lby,Karbstein:2019oej}. We obtain
\begin{equation}
 {\cal I}_{12}(k)\,
 =\,8(2\pi)^{\frac{5}{4}}\Bigl(\frac{\sqrt{\alpha}}{m_e^2}\Bigr)^3  \sqrt{W_1} W_{2} \frac{\sqrt{\tau_1}}{w_2} \frac{w_{0,1}}{w_1} \,
  {\rm e}^{-\frac{1}{2}(\frac{\tau_2\omega}{4})^2(\frac{\tau_2}{T_2})^2\delta\hat{\rm k}^2} \frac{1}{\sqrt{B}}\,{\rm e}^{-\frac{1}{4}(A+\frac{C+D}{B})}\,, \label{eq:I12b}
\end{equation} 
with 
\begin{align}
 A\,=\ &(w_{0,1}\omega)^2\Bigl(\frac{w_{0,1}}{w_1}\Bigr)^2(1+\delta\hat{\rm k})^2\sin^2\varphi\sin^2\vartheta\,, \label{eq:A} \\
 B\,=\ &4(1-\cos\vartheta_{\rm coll})^2+\frac{T_1T_2}{w_1w_2}\sin^2\vartheta_{\rm coll} \,,\label{eq:B} \\
 C\,=\ &4(w_{0,1}\omega)^2\Bigl(\frac{w_{0,1}}{w_1}\Bigr)^2(1+\delta\hat{\rm k})^2h_{\varphi,\vartheta}^2(\sin\vartheta_{\rm coll},1-\cos\vartheta_{\rm coll}) \,,\label{eq:C}
\end{align}
and
\begin{align}
 D\,=\ &\frac{T_1T_2\omega^2}{\sqrt{2}}\biggl\{\Bigl(\frac{w_{0,1}}{w_1}\Bigr)^2\Bigl[(1+\delta\hat{\rm k})(1-\cos\vartheta)-\delta\hat{\rm k}\Bigl(\frac{\tau_2}{T_2}\Bigr)^2(1-\cos\vartheta_{\rm coll})\Bigr]^2 \nonumber\\
 &+\Bigl(\frac{w_{0,2}}{w_2}\Bigr)^2\Bigl[(1+\delta\hat{\rm k})h_{\varphi,\vartheta}(\cos\vartheta_{\rm coll},\sin\vartheta_{\rm coll})
 +\delta\hat{\rm k}\Bigl(\frac{\tau_2}{T_2}\Bigr)^2(1-\cos\vartheta_{\rm coll})\Bigr]^2\biggr\} \,. \label{eq:D}
\end{align}
Here, $\delta\hat{\rm k}=\frac{{\rm k}-\omega}{\omega}$ measures the relative deviation of the signal photon energy from ${\rm k}=\omega$
and the function $h_{\varphi,\vartheta}(a,b)$ in Eqs.~\eqref{eq:C} and \eqref{eq:D} is given by
\begin{equation}
 h_{\varphi,\vartheta}(a,b)=a(1-\cos\vartheta)-b\cos\varphi\sin\vartheta\,. 
\end{equation}
For a relatively compact representation, we have moreover made use of the definitions $T_1=\tau_1\sqrt{1+2(\frac{\tau_1}{\tau_2})^2}$ and $T_2=\tau_2\sqrt{1+\frac{1}{2}(\frac{\tau_2}{\tau_1})^2}$, fulfilling 
$T_1\simeq\tau_1|_{\tau_2\gg\tau_1}$ and $T_2\simeq\tau_2|_{\tau_1\gg\tau_2}$, respectively.
Note also the identities $(\frac{\tau_1}{T_1})^2+(\frac{\tau_2}{T_2})^2=1$ and $\tau_1^2(\frac{\tau_1}{T_1})^2=\frac{\tau_2^2}{2}(\frac{\tau_2}{T_2})^2$.
Analogously, we have introduced $w_1=w_{0,1}\sqrt{1+2(\frac{w_{0,1}}{w_{0,2}})^2}$ and $w_2=w_{0,2}\sqrt{1+\frac{1}{2}(\frac{w_{0,2}}{w_{0,1}})^2}$, for which similar identities hold.
Subsequently, we focus on the case of $\tau_1\geq\tau_2$, which directly implies $0\leq(\frac{\tau_1}{T_1})^2\leq\frac{1}{3}$ and $\frac{2}{3}\leq(\frac{\tau_2}{T_2})^2\leq1$. The reason for concentrating on this regime will become clear below. 

With these results the differential number of polarization-flipped signal photons~\eqref{eq:dNperp} can be expressed as 
\begin{align}
\frac{{\rm d}^3N_\perp}{{\rm dk}{\rm d}\varphi\,{\rm d}\!\cos\vartheta}\,=\ &\frac{4\sqrt{2\pi}}{2025}\Bigl(\frac{\alpha}{\pi}\Bigr)^4
\,(1+\delta{\rm k})^3\,\frac{\omega^3}{m_e^8}\frac{\tau_1W_1 W_2^2}{w_2^2}\Bigl(\frac{w_{0,1}}{w_1}\Bigr)^2\,(1-\cos\vartheta_{\rm coll})^2  \nonumber\\
&\times\Bigl\{ 
\bigl[\cos\varphi
(1-\cos\vartheta\cos\vartheta_{\rm coll})-\sin\vartheta\sin\vartheta_{\rm coll}
\bigr]  f(\beta_1+\beta_2,\beta_\perp+\beta_2) \nonumber\\
&\quad\quad-\sin\varphi(\cos\vartheta-\cos\vartheta_{\rm coll})\, 
g(\beta_1+\beta_2,\beta_\perp+\beta_2)\Bigr\}^2 \nonumber\\
&\times\frac{1}{B}  \,
  {\rm e}^{-(\frac{\tau_2\omega}{4})^2(\frac{\tau_2}{T_2})^2\delta\hat{\rm k}^2
  -\frac{1}{2}(A+\frac{C+D}{B})} \,. \label{eq:dNperpres}
\end{align}
The differential number of signal photons~\eqref{eq:dNperpres} becomes maximal when the exponential suppression is minimized. This is the case for ${\rm k}=\omega$ and $\vartheta=0$, independently of the value of $\vartheta_{\rm coll}$. For $|\delta\hat{\rm k}|>0$ and $\vartheta>0$, it decays rapidly towards zero.
At the same time, the other contributions to \Eqref{eq:I12b}, neglected here from the outset, are exponentially suppressed throughout this parameter regime \cite{Karbstein:2015xra,Karbstein:2016lby}.
The first factor in the exponential in \Eqref{eq:dNperpres} ensures that the signal stems from $\delta\hat{\rm k}\ll1$: for pulse durations $\tau_2\gtrsim3\,{\rm fs}$ and a photon energy of $\omega=9835\,{\rm eV}$ as considered here, we have $(\frac{\tau_2\omega}{4})^2(\frac{\tau_2}{T_2})^2\gtrsim2(6473)^2$.
This implies that the signal receives an overall exponential suppression with $\delta\hat{\rm k}$ independently of the angles; the associated $1/{\rm e}^2$ half-width is $\lesssim2\times10^{-4}$.

For $\vartheta\ll1$, we have $\beta_\perp\simeq\beta_1-\varphi+\frac{\pi}{2}$. In this limit, the square of the term in the curly brackets in \Eqref{eq:dNperpres} simplifies significantly and becomes $\{\cdot\}^2\to\frac{9}{4}(1-\cos\vartheta_{\rm coll})^2\sin^2\bigl(2(\beta_1+\beta_2)\bigr)$.
As this expression encodes the entire dependence of the signal on the polarizations of the pump and probe beams, it implies that the signal becomes maximal for the choice of $\beta_1+\beta_2=\frac{\pi}{4}$, independently of the collision angle. We adopt this optimal choice in the remainder of this article.

The fact that the main contribution to the signal arises from $\delta\hat{\rm k}\ll1$ allows for further simplifications:
when performing the integration over the signal photon energy $\rm k$, we can neglect the subleading corrections of ${\cal O}(\delta\hat{\rm k})$ in the overall prefactor in \Eqref{eq:dNperpres} and formally extend the integration limits of the energy integral to $\pm\infty$, such that $\int{\rm dk}\to\omega\int_{-\infty}^\infty{\rm d}\delta\hat{\rm k}$; see also Refs.~\cite{Karbstein:2018omb,Karbstein:2019oej}.  
With these approximations, we end up with the following closed-form expression for the emission-angle-resolved differential signal photon number
\begin{align}
\frac{{\rm d}^2N_\perp}{{\rm d}\varphi\,{\rm d}\!\cos\vartheta}\,\approx\ &\frac{2\pi}{225}\Bigl(\frac{\alpha}{\pi}\frac{\omega}{m_e}\Bigr)^4(1-\cos\vartheta_{\rm coll})^4\,\frac{W_1 W_2^2}{m_e^3}\frac{\tau_1}{m_ew_2^2}\Bigl(\frac{w_{0,1}}{w_1}\Bigr)^2 \nonumber\\
&\times\frac{1}{B\sqrt{\frac{1}{8}(\tau_2\omega)^2(\frac{\tau_2}{T_2})^2+A_0+\frac{C_0+D_2}{B}}}  \,
  {\rm e}^{-\frac{1}{2}(A_0+\frac{C_0+D_0}{B})}\,{\rm e}^{\frac{\frac{1}{2}(A_0+\frac{C_0+D_1}{B})^2}{\frac{1}{8}(\tau_2\omega)^2(\frac{\tau_2}{T_2})^2+A_0+\frac{C_0+D_2}{B}}} \,, \label{eq:dNperpres_approx}
\end{align}
where the quantities
\begin{align}
A_0\,=\ &A|_{\delta\hat{k}=0}\,, \quad C_0\,=\ C|_{\delta\hat{k}=0} \,, \quad D_0\,=\ D|_{\delta\hat{k}=0}\,, \nonumber\\
D_1\,=\ &\frac{1}{2}\frac{\partial D}{\partial\delta\hat{k}}\Big|_{\delta\hat{k}=0}\,, \quad D_2\,=\ \frac{1}{2}\frac{\partial^2 D}{\partial\delta\hat{k}^2}\Big|_{\delta\hat{k}=0}
\end{align}
are defined in terms of the functions $A$, $C$ and $D$ introduced in Eqs.~\eqref{eq:A}, \eqref{eq:C} and \eqref{eq:D}.
As the terms in the exponential ensure the signal to emerge for small values of $\vartheta\ll1$ only, we can in addition safely approximate all terms in the overall prefactor multiplying the exponential by their leading contribution in the limit of small $\vartheta$.
This results in the following replacement 
\begin{equation}
 \frac{1}{B\sqrt{\frac{1}{8}(\tau_2\omega)^2(\frac{\tau_2}{T_2})^2+A_0+\frac{C_0+D_2}{B}}}\ \to\ \frac{1}{\sqrt{B}}\frac{2}{\tau_1\omega}\frac{T_1}{\tau_1}\frac{1}{\sqrt{B+4\sqrt{2}\,\frac{T_1}{T_2}(1-\cos\vartheta_{\rm coll})^2}} \label{eq:approxfactor}
\end{equation}
of the first term in the second line of \Eqref{eq:dNperpres_approx}. We have explicitly checked numerically that this substitution does not compromise the precision of our explicit results given below.
Due to the fact that $B$ is independent of $\omega$, this substitution makes it manifest that the prefactor in \Eqref{eq:dNperpres_approx} scales cubic with the photon energy $\omega$.

The differential number of perpendicularly polarized signal photons~\eqref{eq:dNperpres_approx} should be compared with $\frac{{\rm d}^2N_1}{{\rm d}\varphi\,{\rm d}\!\cos\vartheta}=\frac{1}{2\pi}\frac{{\rm d}N_1}{{\rm d}\!\cos\vartheta}$, describing the angular decay of the driving laser photons~\eqref{eq:decaydriver} of beam $b=1$ traversing the interaction region unaltered. The dependence of $N_1$ on the azimuthal angle $\varphi$ is trivial as the driving laser field exhibits a rotational symmetry about its beam axis.

An important criterion discriminating if the polarization-flipped signal photons emitted in $(\varphi,\vartheta)$ direction are background dominated or not is to assess if the ratio of $\frac{{\rm d}^2N_\perp}{{\rm d}\varphi\,{\rm d}\!\cos\vartheta}$ and $\frac{{\rm d}^2N_1}{{\rm d}\varphi\,{\rm d}\!\cos\vartheta}$ surpasses the polarization purity $\cal P$ of the employed polarimeter. We call the signal photons which can be measured above the background of the driving laser photons {\it discernible}. Discernible signal photons fulfill the criterion \cite{Karbstein:2016lby}
\begin{equation}
 \frac{{\rm d}^2N_\perp}{{\rm d}\varphi\,{\rm d}\!\cos\vartheta}\geq {\cal P}\frac{{\rm d}^2N_1}{{\rm d}\varphi\,{\rm d}\!\cos\vartheta} \,, \label{eq:discern}
\end{equation}
or explicitly, accounting for Eqs.~\eqref{eq:dNperpres_approx} and \eqref{eq:approxfactor},
\begin{align}
&2\ln\biggl\{\frac{1}{\cal P}\frac{8\pi^2}{225}\Bigl(\frac{\alpha}{\pi}\Bigr)^4\Bigl(\frac{\omega}{m_e}\frac{W_2}{m_e}\frac{1}{w_1w_2m_e^2}\Bigr)^2\frac{T_1}{\tau_1}  \,\frac{1}{\sqrt{B}}\frac{(1-\cos\vartheta_{\rm coll})^4}{\sqrt{B+4\sqrt{2}\,\frac{T_1}{T_2}(1-\cos\vartheta_{\rm coll})^2}}\biggl\} \nonumber\\
&\, + (\omega w_{0,1}\vartheta)^2-A_0-\frac{C_0+D_0}{B}+\frac{(A_0+\frac{C_0+D_1}{B})^2}{\frac{1}{8}(\tau_2\omega)^2(\frac{\tau_2}{T_2})^2+A_0+\frac{C_0+D_2}{B}}\geq0 \,. \label{eq:P}
\end{align}
We denote the number of perpendicularly polarized, discernible signal photons by $N_{\perp,{\rm dis}}$.
Apart from the polarization purity, the common photon energy and the collision angle, this expression depends on the duration and waist of the probe beam $b=1$, as well as the pulse energy, duration and waist of the pump beam $b=2$.
For each value of the azimuthal angle $\varphi$, it can be solved numerically for the polar angles $\vartheta=\vartheta(\varphi)$ fulfilling the discernibility criterion.

\subsection{Analytical scalings}\label{sec:scalings}

Before evaluating these formulae for experimentally feasible parameters, let us derive analytical estimates allowing for a simple interpretation of the results.
Using that the signal photons are predominantly emitted in the forward direction with $\vartheta\ll1$, we expand the exponential in \Eqref{eq:dNperpres_approx} and \Eqref{eq:P} to leading order $\sim\vartheta^2$ and neglect higher-order contributions.
Following Refs.~\cite{Karbstein:2018omb,Karbstein:2019oej}, this allows for closed-form approximations of the attainable signal photon numbers.
Of course, this approximation may not quantitatively reproduce the double differential signal photon number $\frac{{\rm d}^2N_\perp}{{\rm d}\varphi\,{\rm d}\!\cos\vartheta}$ for very small scattering centers resulting in a comparably wide scattering range.
In this case, higher-order terms in $\vartheta$ may be essential to reproduce the angular decay of the signal photon distribution towards larger values of $\vartheta$ accurately.
However, even in this case it should still allow for qualitative insights into the scaling of the single differential signal photon number $\frac{{\rm d}N_\perp}{{\rm d}\varphi}$ with various parameters; cf. below. 
The corresponding results are
\begin{align}
 \frac{{\rm d}N_\perp}{{\rm d}\varphi}&\simeq\frac{4\pi}{225}\Bigl(\frac{\alpha}{\pi}\Bigr)^4\frac{\omega}{m_e}(1-\cos\vartheta_{\rm coll})^4\frac{W_1W_2^2}{m_e^3}\frac{\lambdabar_{\rm C}^4}{(w_1w_2)^2}\frac{T_1}{\tau_1}\sqrt{\frac{\cal B}{B}} \nonumber\\
 &\quad\times\frac{1}{(\frac{w_{0,1}}{w_1})^2{\cal B}+\frac{T_1T_2}{w_1w_2}(\frac{w_{0,2}}{w_2})^2\sin^2\vartheta_{\rm coll}\cos^2\varphi}\,, \label{eq:dNperpdphiapprox}
\end{align}
and
\begin{align}
 \frac{{\rm d}N_{\perp,{\rm dis}}}{{\rm d}\varphi}&\simeq\frac{\cal P}{2\pi}\frac{W_1}{\omega} \frac{{\cal B}}{(\frac{w_{0,1}}{w_1})^2{\cal B}+\frac{T_1T_2}{w_1w_2}(\frac{w_{0,2}}{w_2})^2\sin^2\vartheta_{\rm coll}\cos^2\varphi}\nonumber\\
 &\quad\times\biggl\{\frac{2\pi}{15}\Bigl(\frac{\alpha}{\pi}\Bigr)^2 \frac{\omega}{m_e}\frac{W_2}{m_e} (1-\cos\vartheta_{\rm coll})^2\frac{\lambdabar_{\rm C}^2}{w_1w_2}\sqrt{\frac{2}{\cal P}\frac{T_1}{\tau_1}}\frac{1}{(B{\cal B})^{1/4}}\biggr\}^\kappa\,, \label{eq:dNperpdisdphiapprox}
\end{align}
where $\lambdabar_{\rm C}=\frac{1}{m_e}$ is the reduced Compton wavelength of the electron. In addition to the expression for $B$ defined in \Eqref{eq:B}, we have introduced the shorthand notations
\begin{align}
 B&=4(1-\cos\vartheta_{\rm coll})^2+\frac{T_1T_2}{w_1w_2}\sin^2\vartheta_{\rm coll}\,, \nonumber\\
 {\cal B}&=B+4\sqrt{2}\,\frac{T_1}{T_2}(1-\cos\vartheta_{\rm coll})^2\,, \nonumber\\
 \kappa&=2\Bigl(\frac{w_2}{w_{0,2}}\Bigr)^2\frac{{\cal B}}{{\cal B}-\frac{T_1T_2}{w_1w_2}\sin^2\vartheta_{\rm coll}\cos^2\varphi}\,. \label{eq:calB,kappa}
\end{align}
Equations~\eqref{eq:dNperpdphiapprox} and \eqref{eq:dNperpdisdphiapprox} suggest that the total number of polarization-flipped signal photons $N_\perp$ should scale differently with the physical parameters characterizing the driving beam configuration than the number of discernible signal photons $N_{\perp,{\rm dis}}$, constituting a subset of the former. This is in line with previous findings for head-on colliding beams \cite{Karbstein:2018omb,Karbstein:2019oej}. However, due to the additional parameter of $\vartheta_ {\rm coll}$ here the phenomenology is richer.
Of particular interest are the parameter regimes where the denominators in Eqs.~\eqref{eq:dNperpdphiapprox} and \eqref{eq:dNperpdisdphiapprox} are dominated either by the contribution $\sim\frac{T_1T_2}{w_1w_2}\sin^2\vartheta_{\rm coll}$ or by $\sim(1-\cos\vartheta_{\rm coll})^2$ for all values of $\varphi$.
Assuming the waists of the two beams to be  of the same order, these two regimes are characterized by the condition $\frac{T_1T_2}{w_1w_2}\sin^2\vartheta_{\rm coll}\gg4(1+\sqrt{2}\frac{T_1}{\tau_1})(1-\cos\vartheta_{\rm coll})^2$ or by its complement.
First, we consider the scaling of the signal photon numbers with the beam waist. Here, we infer $N_\perp\sim(\frac{1}{w_{0,1}w_{0,2}})^n$ and $N_{\perp,{\rm dis}}\sim(\frac{1}{w_{0,1}w_{0,2}})^{\frac{n}{2}[2+(w_{0,2}/w_{0,1})^2]}$, with $n=1$ in the former and $n=2$ in the latter regime.
Similarly, we extract the following scaling with the pulse durations: $N_\perp\sim(\frac{1}{\tau_1\tau_2})^{2-n}$ and $N_{\perp,{\rm dis}}\sim(\frac{1}{\tau_1\tau_2})^{\frac{2-n}{2}[2+(w_{0,2}/w_{0,1})^2]}$.
Besides, the signal photon numbers scale with the pulse energies as $N_\perp\sim W_1W_2^2$ and $N_{\perp,{\rm dis}}\sim W_1W_2^{2+(w_{0,2}/w_{0,1})^2}$. Interestingly, the last behavior is independent of $n$ and thus the same in both regimes.

Lastly, we note that \Eqref{eq:dNperpdphiapprox} even permits an analytical integration over the azimuthal angle $\varphi$. In turn, the total number of polarization-flipped signal photons normalized by the incident number of photons for probing $N\simeq\frac{W_1}{\omega}$ is given by
\begin{align}
 \frac{N_\perp}{N}&\simeq\,\frac{8\pi^2}{225}\Bigl(\frac{\alpha}{\pi}\Bigr)^4\Bigl(\frac{\omega}{m_e}\frac{W_2}{m_e}\Bigr)^2\frac{\lambdabar_{\rm C}^4}{(w_1w_2)^2}\frac{T_1}{\tau_1}\frac{w_1}{w_{0,1}}\frac{(1-\cos\vartheta_{\rm coll})^4}{\sqrt{4(1-\cos\vartheta_{\rm coll})^2+\frac{T_1T_2}{w_1w_2}\sin^2\vartheta_{\rm coll}}} \nonumber\\
 &\quad\times\frac{1}{\sqrt{4(\frac{T_1}{\tau_1}\frac{w_{0,1}}{w_1})^2(1-\cos\vartheta_{\rm coll})^2+\frac{T_1T_2}{w_1w_2}\sin^2\vartheta_{\rm coll}}}\,. \label{eq:Nperpapprox}
\end{align}
As expected, this result agrees for $\vartheta_{\rm coll}=\pi$ with the one determined in Ref.~\cite{Karbstein:2018omb} for counter-propagating laser pulses at zero impact in the limit of infinitely long Rayleigh lengths; cf. Eqs. (7) and (11) of \cite{Karbstein:2018omb}.
In particular, the number of polarization-flipped signal photons~\eqref{eq:Nperpapprox} in this limit becomes independent of both the pump and probe pulse durations.
Conversely, for any finite collision angle, this number depends explicitly on each of these time scales, resulting in distinctly different scalings of the signal photon numbers with the various parameters of the driving laser pulses; see above.

\section{Experimental Parameters and Results}
\label{sec:setup}

\begin{figure}[b]
 \includegraphics[width=0.75\linewidth]{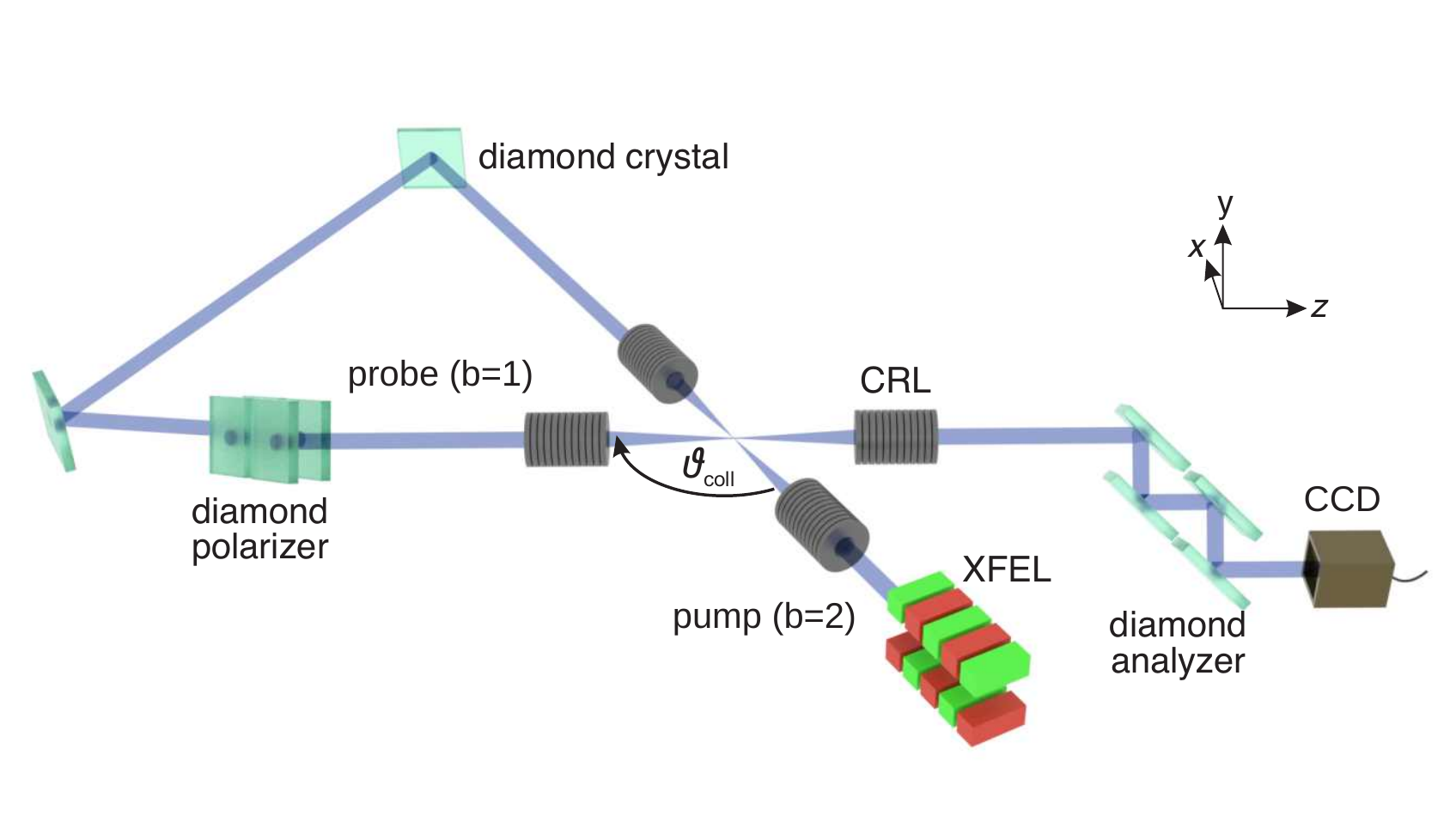}\vspace{-5mm}
 \caption{Illustration of the experimental setup utilizing compound refractive lenses (CRLs) to focus and re-collimate the XFEL beam. Reflections at diamond crystals change the propagation direction, and a pair of diamond quasi-channel-cuts serve as polarizer and analyzer, respectively. The original XFEL beam is focused with a CRL to constitute the pump field; the beam focus defines the interaction point. Subsequently, it is defocused with a CRL and by reflection at two diamond crystals directed back to the interaction point under an angle of $\vartheta_{\rm coll}$. Before reaching the interaction point, it is polarized with a diamond polarizer and the resulting probe beam focused to the interaction point with a CRL. Finally, it is defocused with another CRL, analyzed with a diamond analyzer and the signal registered with a CCD.}
 \label{fig:setup}
\end{figure}

For the remainder, we assume a photon energy of $\omega=9835\,{\rm eV}$ for the incident XFEL pulse and a repetition rate of $27\,000\,{\rm pulses}/{\rm s}$ \cite{XFEL}. Precisely for this energy, the x-ray polarimeter consisting of a pair of diamond quasi-channel-cuts utilizing four 400 Bragg reflections with $\theta=45^\circ$, forming an integral part of our setup (cf. Fig.~\ref{fig:setup}),
was shown to achieve a polarization purity of ${\cal P}=1.4\times10^{-10}$ \cite{Bernhardt:2020vxa}.
Due to its high heat conductivity and its small thermal expansion, Diamond is the preferred choice for applications with high-repetition rate and the corresponding high heat load.  
Moreover, we consider the following options for the pulse duration $\tau^{\rm HM}$ at half-maximum and the associated pulse energy $W(\tau)$ of the XFEL pulse at this photon energy,
\begin{table}[h!]
\begin{tabular}{|c||c|c|}
\hline
 Option \# & $\tau^{\rm HM}\,[{\rm fs}]$ & $W\,[{\rm mJ}]$ \\
 \hline
 \hline
 1 & 1.7 & 0.072 \\
 2 & 23 & 0.85 \\
 3 & 107 & 2.4 \\
\hline
 4 & 14 & 11 \\
\hline
\end{tabular}\,.
\end{table}

\noindent Options 1-3 amount to parameters currently available at the European XFEL \cite{XFEL}: to obtain the pulse energies listed here, we have interpolated the values given for photon energies $12.4\,{\rm keV}$ and $8.27\,{\rm keV}$ in Tables C.5 and C.6 of Ref.~\cite{Schneidmiller:92341} attainable with a $17.5\,{\rm GeV}$ electron beam.
Besides, Option 4 serves as an example of idealized, theoretically projected parameters based on state-of-the-art technology \cite{Chubar:2015mpa}.
Moreover, we emphasize that the pulse durations of all options are such that the infinite Rayleigh range approximation allows for reliable insights at arbitrary collision angles at least for $w_{0,b}^{\rm HM}\gtrsim50\,{\rm nm}$; cf. Sec.~\ref{sec:infiniteRayleigh}.

The basic setup envisioned by us for studying vacuum birefringence in a purely XFEL-based experiment is depicted in Fig.~\ref{fig:setup}. At the European XFEL, the temporal separation between subsequent x-ray pulses is for instance given by $222\,{\rm ns}$. This implies that in order to collide two consecutive pulses, the optical path separating the beam foci needs to be chosen as $66.6\,{\rm m}$.
For the realization of specific collision angles in experiment, see below.

\subsection{Idealized assumptions}\label{subsec:Idealized}

In a next step, we analyze the numbers of polarization-flipped signal photons attainable with the setup depicted in Fig.~\ref{fig:setup} -- for the moment ignoring any non-trivial losses and pulse deformations induced by the optics used to guide and polarize the beam.
The latter will be accounted for in Sec.~\ref{subsec:Realistic} below.
Being mainly interested in the qualitative behavior of the attainable signal, we base all our considerations in the present section on the analytical estimates derived in Sec.~\ref{sec:scalings}.

Correspondingly, we have $\tau_1=\tau_2=\tau$.
To rotate the polarization direction of the probe beam $b=1$ by $\frac{\pi}{4}\,{\rm rad}$ relatively to that of the pump beam $b=2$, the quasi-channel cut acting as polarizer has to be rotated by an angle of $45^\circ$ and thus only accepts half of the incident photons, resulting in a loss of $50\%$.
This results in $2W_1=W_2=W$.
Moreover, we consider two different scenarios for the focusing, namely $w_{0,1}^{\rm HM}=w_{0,2}^{\rm HM}=w_0^{\rm HM}=50\,{\rm nm}$ and $5\,{\rm nm}$, respectively.
As noted above, particularly the first value amounts to an experimentally viable option. The latter one is more ambitious, but should be feasible in principle \cite{Mimura:2014,Yamauchi:2011,Huang:2013}.
For these parameters, \Eqref{eq:Nperpapprox} assumes a simple form, 
\begin{align}
 \frac{N_\perp}{N}\simeq\,\frac{16\pi^2}{675}\Bigl(\frac{\alpha}{\pi}\frac{\lambdabar_{\rm C}}{w_0}\Bigr)^4\Bigl(\frac{\omega}{m_e}\frac{W}{m_e}\Bigr)^2c(\tfrac{\tau}{4w_0},\vartheta_{\rm coll}) \,, \label{eq:NperpbyNequalparams}
\end{align}
where we have made use of the fact that $N\simeq\frac{W}{2\omega}$ and introduced the function
\begin{equation}
 c(\tfrac{\tau}{4w_0},\vartheta_{\rm coll})=\frac{\frac{1}{4}(1-\cos\vartheta_{\rm coll})^2}{1+(\frac{\tau}{4w_0}\frac{2\sin\vartheta_{\rm coll}}{1-\cos\vartheta_{\rm coll}})^2}. \label{eq:c}
\end{equation}
This function parameterizes the deviations from the formal limit of $\vartheta_{\rm coll}\to\pi$ for which the dependence of the pulse duration drops out completely and \Eqref{eq:NperpbyNequalparams} is maximized; note that $c(\tfrac{\tau}{4w_0},\pi)=1$.

\begin{figure}
 \includegraphics[width=1\linewidth]{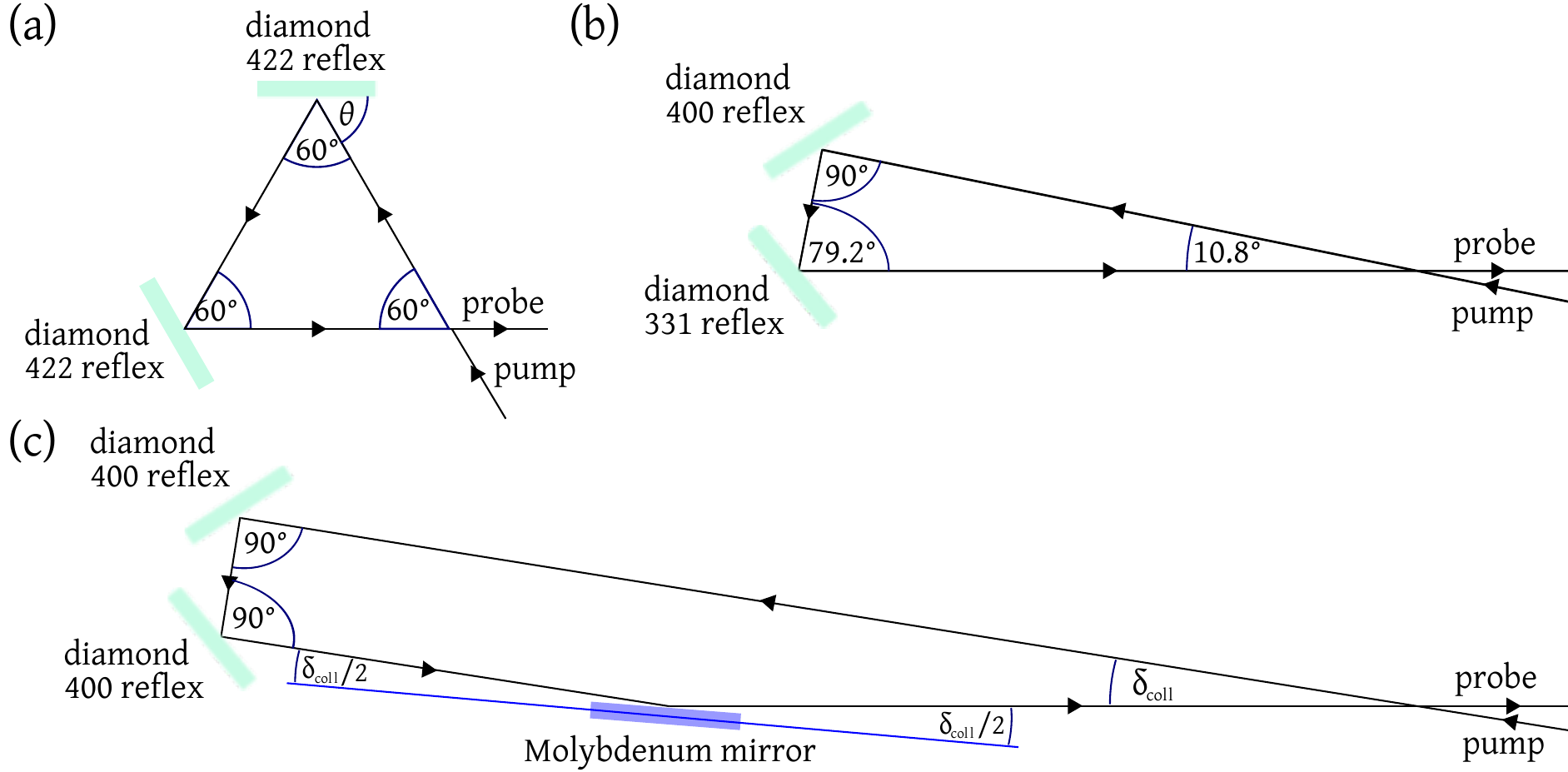}
 \caption{Beam guidance in the three Setups (a)-(c) analyzed in the present work, highlighting the relevant diamond reflexes envisioned to change the propagation direction of the XFEL beam. In Setup (a) we exemplarily indicate the reflection angle $\theta$ at the first diamond crystal. We only employ symmetric reflections, such that the incident reflection angle always equals the outgoing one. Apart from diamond crystals, Setup (c) also involves a Molybdenum mirror allowing for variable choices of the collision angle $\vartheta_{\rm coll}=180^\circ-\delta_{\rm coll}$ with $\delta_{\rm coll}\gtrsim0.3^\circ$.}
 \label{fig:possibilities}
\end{figure}

Equation~\eqref{eq:NperpbyNequalparams} clearly highlights that the decay of the number of polarization-flipped signal photons for generic collision angles is not just governed by a trigonometric function, but in fact controlled by the dimensionless ratio of pulse duration and waist diameter $\frac{\tau}{4w_0}$.
The larger this ratio, the faster the decay of $N_\perp$ as a function of the angle $\delta_{\rm coll}=\pi-\vartheta_{\rm coll}$ measuring the deviation from $\vartheta_{\rm coll}\to\pi$ maximizing the signal photon yield; cf. also Fig.~\ref{fig:possibilities}.
Here, we focus on the scaling of $N_\perp/N$ for small deviations from the counter-propagation geometry:
given that the condition $\delta_{\rm coll}\ll1$ is fulfilled, we can identify two extremal cases characterized by either $\frac{\tau}{4w_0}\delta_{\rm coll}\ll1$ or $\frac{\tau}{4w_0}\delta_{\rm coll}\gg1$, respectively.
From \Eqref{eq:c}, we infer the following scalings
\begin{equation}
 c(\tfrac{\tau}{4w_0},\vartheta_{\rm coll})\simeq
 \left\{\begin{array}{c}
        1 \\ {[\frac{\tau}{4w_0}\delta_{\rm coll}]^{-2}}
        \end{array}\right\}\quad\text{for}\quad
        \begin{array}{c}
         \frac{\tau}{4w_0}\delta_{\rm coll}\ll 1 \\ \frac{\tau}{4w_0}\delta_{\rm coll}\gg 1
        \end{array}\quad\text{and}\quad \delta_{\rm coll}\ll1\,. \label{eq:NperpbyNequalparamslimits}
\end{equation}
These findings immediately imply that the harder the focusing, the larger the impact of a finite value of $\delta_{\rm coll}$ on the signal. Due to their small wavelength, for x-rays the ratio $\frac{\tau}{4w_0}$ can be much larger than for optical beams. 
For the different pulse durations of Options 1-4 and $w_0^{\rm HW}=50\,{\rm nm}$ ($5\,{\rm nm}$) this ratio ranges from $\frac{\tau}{4w_0}\simeq2.5$ ($25$) for the minimum pulse duration of Option 1 to a value as large as $\frac{\tau}{4w_0}\simeq160$ ($1604$) for the maximum value of Option 3.

Particularly viable choices for the collision angle in an actual experiment are (a): $\vartheta_{\rm coll}=120^\circ$, (b): $\vartheta_{\rm coll}=169.2^\circ$ and (c): $\vartheta_{\rm coll}=179.7^\circ$ upon fixing $\delta_{\rm coll}=0.3$.
These values of $\vartheta_{\rm coll}$ can be implemented by using specific reflexes at diamond to change the propagation direction of the probe relatively to the pump. For a graphical representation of the different Setups (a)-(c) considered by us, see Fig.~\ref{fig:possibilities}.
In Setup (a), we envision the use of two $422$ reflexes (extinction depth $\Lambda=4.8\,\upmu{\rm m}$) with a reflection angle of $\theta=60^\circ$. Conversely, Setup (b) is based on two different reflexes: a $400$ reflex ($\Lambda=3.6\,\upmu{\rm m}$) coming along with $\theta=45^\circ$, and a $331$ reflex ($\Lambda=5.7\,\upmu{\rm m}$) implying $\theta=50.4^\circ$.
Finally, Setup (c) uses two $400$ reflexes in combination with a molybdenum mirror (roughness $0.03\,{\rm nm}$).

For Setup (a) the suppression factor $c(\tfrac{\tau}{4w_0},\vartheta_{\rm coll})$ defined in \Eqref{eq:c} is as small as $0.06$ ($7\times10^{-4}$) for the minimum pulse duration of Option 1, but becomes even smaller, namely $1.6\times10^{-5}$ ($1.6\times10^{-7}$) for the maximum pulse duration of Option 3.
Analogously, for Setup (b) the suppression factor is $0.8$ ($0.04$) for Option 1 and $0.001$ ($1.1\times10^{-5}$) for Option 3.
Finally, the values for Setup (c) and the minimum value of $\delta_{\rm coll}\simeq0.3^\circ$ achievable in experiment are $0.99$ ($0.98$) for Option 1, and $0.59$ ($0.01$) for Option 3.

In the next step, we determine both the total number $N_\perp$ of polarization-flipped signal photons and the number of discernible signal photons $N_{\perp,{\rm dis}}$ from Eqs.~\eqref{eq:dNperpres_approx}-\eqref{eq:P} for Options 1-4 implemented in the different Setups (a)-(c).
The corresponding results are assembled in Tab.~\ref{tab:results_idealized}.
\begin{table}[htbp]
	\centering
	\begin{tabular}{|c|c||*{6}{c|}}
		\hline
		\multirow{2}{*}{Option \#} & \multirow{2}{*}{$w_0^{\rm HM}\,[{\rm nm}]$}  & \multicolumn{2}{c|}{(a)} & \multicolumn{2}{c|}{(b)} & \multicolumn{2}{c|}{(c): $\vartheta_{\rm coll}=179.7^\circ$}\\
		\cline{3-4} \cline{5-6} \cline{7-8}
        & &\multicolumn{1}{c|}{$N_\perp$/h} & \multicolumn{1}{c|}{$N_{\perp,{\rm dis}}$/h} & \multicolumn{1}{c|}{$N_\perp$/h} & \multicolumn{1}{c|}{$N_{\perp,{\rm dis}}$/h} & \multicolumn{1}{c|}{$N_\perp$/h} & \multicolumn{1}{c|}{$N_{\perp,{\rm dis}}$/h}\\
		\hline
		\hline
        1 & \multirow{3}{*}{50} & $0.1$ & $3.1 \times 10^{-7}$ & $1.5$ & $3.4 \times 10^{-5}$ & $1.8$ & $8.2\times10^{-5}$ \\
		2 & & $1.1$ & $2.8 \times 10^{-6}$ & $68.2$ & $1.6 \times 10^{-3}$ & $2933$ & $1.4$ \\
		3 & & $1.1$ & $1.7 \times 10^{-6}$ & $72.5$ & $9.9 \times 10^{-4}$ & $4.0\times10^5$ & $20.1$ \\
		\hline
		1 & \multirow{3}{*}{5} & $11.9$ & $4.0 \times 10^{-4}$ & $744.4$ & $0.22$  & $1.8\times10^{4}$ & $42.6$ \\
		2 & & $107.1$ & $3.1 \times 10^{-3}$ & $6977$ & $1.8$  &  $6.1\times10^6$ &  $9.6\times10^5$ \\
		3 & & $111.4$ & $1.9 \times 10^{-3}$ & $7259$ & $1.1$  & $8.5\times10^{6}$ & $2.3\times10^{6}$ \\
		\hline
		\multirow{2}{*}{4} & 50 & $6255$ & $0.4$ & $3.8\times10^5$ & $221.6$  & $6.5\times10^{6}$ & $42955$ \\
		& 5  & $6.3\times10^5$ & $468.5$ & $4.1 \times 10^{7}$ & $3.0\times10^5$  & $2.5\times10^{10}$ & $6.4\times10^{9}$ \\
		\hline
	\end{tabular}
	\caption{Results for the numbers of signal photons per hour based on the idealized assumptions in Sec.~\ref{subsec:Idealized}.
	For each Setup (a)-(c) we provide results for the different Options 1-4 characterized by different XFEL pulse durations and energies.}
	\label{tab:results_idealized}%
\end{table}	

For the majority of the cases considered here, the parameters of Setups (a) and (b) fulfill the condition $\frac{T_1T_2}{w_1w_2}\sin^2\vartheta_{\rm coll}\gg4(1+\sqrt{2}\frac{T_1}{\tau_1})(1-\cos\vartheta_{\rm coll})^2$ introduced below \Eqref{eq:calB,kappa}; the only exception is Setup (b) with $w_0^{\rm HM}=50\,{\rm nm}$ and Option 1.
Indeed, -- aside from this exception -- the results of a full numerical evaluation of Eqs.~\eqref{eq:dNperpres_approx}-\eqref{eq:approxfactor} for Setups (a) and (b) listed in Tab~\ref{tab:results_idealized} roughly fulfill $N_\perp|_{w_0^{\rm HM}=5\,{\rm nm}}\approx10^2\,N_\perp|_{w_0^{\rm HM}=50\,{\rm nm}}$ and $N_{\perp,{\rm dis}}|_{w_0^{\rm HM}=5\,{\rm nm}}\approx10^3\,N_{\perp,{\rm dis}}|_{w_0^{\rm HM}=50\,{\rm nm}}$ for fixed other parameters.
This behavior is in line with the analytical scalings estimated in Sec.~\ref{sec:scalings} for the present parameter regime.
The scalings for Setup (b) and Option 1 are different as the case with $w_0^{\rm HM}=50\,{\rm nm}$
belongs to the complementary parameter regime.
In this case, the increase of the signal photon number  when reducing the beam waist from $w_0^{\rm HM}=50\,{\rm nm}$ to $w_0^{\rm HM}=5\,{\rm nm}$ is more pronounced.

The considerations in Sec.~\ref{sec:scalings} also predict the scalings $N_\perp\sim W(\frac{W}{\tau})^2$ and $N_{\perp,{\rm dis}}\sim W(\frac{W}{\tau})^3$ with regard to both the pulse energy and duration for Setups (a) and (b).
Plugging the parameters of Options 1-4 into these relations, we find that the attainable signal photon numbers should be ordered as $N_\perp|_{{\rm Option}\,1}<N_\perp|_{{\rm Option}\,2}\lesssim N_\perp|_{{\rm Option}\,3}<N_\perp|_{{\rm Option}\,4}$, while $N_{\perp,{\rm dis}}|_{{\rm Option}\,1}<N_{\perp,{\rm dis}}|_{{\rm Option}\,3}< N_{\perp,{\rm dis}}|_{{\rm Option}\,2}<N_{\perp,{\rm dis}}|_{{\rm Option}\,4}$ for fixed other parameters. Also this behavior is clearly reflected by our numerical results for Setups (a) and (b) in Tab.~\ref{tab:results_idealized}. 

For Setup (c) the situation is less transparent, because some of the parameter sets  meet the condition $\frac{T_1T_2}{w_1w_2}\sin^2\vartheta_{\rm coll}\gg4(1+\sqrt{2}\frac{T_1}{\tau_1})(1-\cos\vartheta_{\rm coll})^2$ and several others its complement. 

Apart from those associated with the idealized, theoretically projected Option $4$, for Setups (a) and (b) the prospective numbers of discernible signal photons per hour are rather small.
Despite the high repetition rate, they are in particular much smaller than those attainable in the conventional collision scenario envisioning the head-on collision of an XFEL and a petawatt class optical high-intensity laser beam \cite{Karbstein:2018omb}.
However, an important asset of the XFEL-only scenario studied here is the inherent synchronization of the two colliding pulses.
On the other hand, the signal photon numbers accessible with Setup (c) -- in particular those for $w_0^{\rm HM}=5\,{\rm nm}$ -- look quite promising.
In the next section we aim at assessing if the observed tendencies persist when a more refined and accurate description of the collision scenarios sketched in Figs.~\ref{fig:setup} and \ref{fig:possibilities} is invoked.

\subsection{Refined Modeling}\label{subsec:Realistic}

In an actual experiment, the pulse duration of the probe $\tau_1$ typically deviates from the pulse duration of the pump $\tau_2=\tau$. In the present case, we have $\tau_1\geq\tau_2$. The reason for this are the reflections at the two diamond crystals used to change its propagation direction, as well as the additional four reflections of the original XFEL pulse within the quasi-channel-cut polarizer; see Fig.~\ref{fig:setup}.
The increase of the original pulse duration by the reflections can be estimated along the lines of Refs.~\cite{Lindberg:2012tn,Shvydko:2012rzc}.

More specifically, the response function $r_1(t)$ characterizing the deformation of an incident $\delta$-pulse for the electric field after a single reflection is 
\begin{equation}
 r_1(t)=\frac{J_1(t/\tau_r)}{t/\tau_r}\,\Theta(t)\,, \label{eq:r1}
\end{equation}
where $J_1(.)$ denotes the Bessel function of the first kind and $\Theta(.)$ is the Heaviside function.
The reference time scale $\tau_r$ is given by $\tau_r=(2\Lambda/c)\sin\theta$, and hence is fully determined by the extinction depth $\Lambda$ of the particular crystal reflex used to reflect the beam with a reflection angle $\theta$.
Equation~\eqref{eq:r1} implies that the response function associated with $n$ subsequent reflections of the same kind is given by
\begin{equation}
 r_n(t)=\int_0^t {\rm d}T\,r_1(t-T)r_{n-1}(T)\quad\text{for}\quad n\geq2\,.
\end{equation}
Correspondingly, an incident temporal pulse profile $f(t)$ is modified to
\begin{equation}
 f_n(t)=\int_0^\infty{\rm d}T\,r_n(T)f(t-T)
\end{equation}
after $n$ similar reflections.
By iteration, this approach can be straightforwardly adopted to reflections off different crystal materials coming along with individual extinction depths and reflection angles.

In the scenario considered by us, the actual probe pulse duration $\tau_1$ needs to account for the deformations of the initial pulse profile of duration $\tau$ by both the reflections at the two diamond crystals guiding it to the interaction region and the 4 reflections in the diamond polarizer.
The initial pulse profile is $f_{\rm in}(t)={\rm e}^{-4(t/\tau)^2}$; cf. \Eqref{eq:E}.
We find that the resulting reflected pulse $f_{\rm out}(t)$ always consists of a pronounced main peak which can be well-approximated as Gaussian peaked at a temporal offset $t_0$ with several side lobes of much lower amplitude. 
In turn, we identify its pulse duration with the pulse duration $\tau'$ of the approximating Gaussian intensity profile and identify $\tau_1=\tau'$.

The reflection-mediated increase of the pulse duration $\tau\to\tau'$ is generically also accompanied by an effective intensity loss due to redistribution of laser intensity from the main peak into side lobes.
In order to account for this effect we estimate the associated transmission factor as
\begin{equation}
    \mathfrak{t}_{\tau\to\tau'}=\frac{\int_{-\tau'^{\rm HM}/2}^{\tau'^{\rm HM}/2} |f_{\rm out}(t-t_0)|^2}{\int_{-\infty}^\infty |f_{\rm out}(t)|^2}\biggl(\frac{\int_{-\tau^{\rm HM}/2}^{\tau^{\rm HM}/2} |f_{\rm in}(t)|^2}{\int_{-\infty}^\infty |f_{\rm in}(t)|^2}\biggr)^{-1}\,,
\end{equation}
corresponding to the ratio of the intensity fractions contained within the half-maximum pulse durations before and after the deformation of the pulse profile.   

Implementing the steps outlined above, for the three different Setups (a)-(c) considered here (see Fig.~\ref{fig:possibilities}) and the different input pulse durations of Options 1-4 we obtain
\begin{table}[h!]
\begin{tabular}{|c||c|c|c|}
\hline
 $\tau^{\rm HM}\,[{\rm fs}]$ & $\tau_1^{\rm HM}({\rm a})\,[{\rm fs}]$ &  $\tau_1^{\rm HM}({\rm b})\,[{\rm fs}]$ & $\tau_1^{\rm HM}({\rm c})\,[{\rm fs}]$ \\
 \hline
 \hline
 1.7 & 63 & 57 & 50 \\
 14 & 64 & 58 & 51 \\
 23 & 65 & 59 & 52 \\
 107 & 111 & 110 & 108\\
\hline
\end{tabular}\,.
\end{table}

\noindent We note that the use of additional reflection would essentially not change these output pulse durations: we find that the output pulse durations stabilize after about four reflections.

Moreover, for the associated transmission factors we find
\begin{table}[h!]
\begin{tabular}{|c||c|c|c|}
\hline
 $\tau^{\rm HM}\,[{\rm fs}]$ & $\mathfrak{t}_{\tau\to\tau'}({\rm a})$ &  $\mathfrak{t}_{\tau\to\tau'}({\rm b})$ & $\mathfrak{t}_{\tau\to\tau'}({\rm c})$ \\
 \hline
 \hline
 1.7 & 0.96 & 0.98 & 0.84 \\
 14 & 0.96 & 0.98 & 0.86 \\
 23 & 0.96 & 0.98 & 0.89 \\
 107 & 1.00 & 1.00 & 1.00\\
\hline
\end{tabular}\,.
\end{table}

Lastly, we estimate the losses of the initial XFEL pulse when traversing the setup depicted in Figs.~\ref{fig:setup} and \ref{fig:possibilities}.
Each passage through a compound refractive lens (CRL) comes with a loss of the incident energy. This loss depends on the envisioned focusing: for a lens focusing down to a waist of the order of $50\,{\rm nm}$ we estimate a loss of $80\%$, while for a $5\,{\rm nm}$ focus the loss is expected to becomes as large as $98\,\%$ \cite{Huang:2013}.
Besides, each reflection at a diamond surface induces a loss of $2\%$.
In addition, reflections at diamond are characterized by a finite bandwidth of $\Delta\omega_{\rm diamond}=21\,{\rm meV}$. Estimating the bandwidth of the initial XFEL pulses from the time-bandwidth product of nearly Fourier-limited XFEL pulses, $\tau^{\rm HM}\times\Delta\omega\simeq8.0$ \cite{Geloni:2011cu}, we find that the first reflection at a diamond surface reduces the incident XFEL pulse energy by a pulse-duration-dependent factor of $\mathfrak{t}_{\Delta\omega}(\tau)=\Delta\omega/\Delta\omega_{\rm diamond}$.

For the pump pulse durations considered here, we find
\begin{table}[h!]
\begin{tabular}{|c||c|}
\hline
 $\tau^{\rm HM}\,[{\rm fs}]$ & $\mathfrak{t}_{\Delta\omega}$ \\
 \hline
 \hline
 1.7 & $\frac{21}{3096}\simeq 0.0068$   \\[1mm]
 14 & $\frac{21}{376}\simeq 0.056$  \\[1mm]
 23 & $\frac{21}{229}\simeq 0.092$  \\[1mm]
 107 & $\frac{21}{49}\simeq 0.43$  \\[1mm]
\hline
\end{tabular}\,.
\end{table}

\noindent These values are independent of the details of the considered collision setup.
Finally, also the reflection at the Molybdenum mirror employed in Setup (c) comes with a loss: for the minimal reflection angle of $\delta_{\rm coll}=0.3^\circ$ ($\vartheta_{\rm coll}=179.7^\circ$) which can be realized in experiment, its reflectivity is $r\simeq96\%$.

Accounting for all these factors, as well as the factor of $0.5$ for the probe pulse introduced in Sec.~\ref{subsec:Idealized} above, the energies of the pump and probe pulses can be estimated as (cf. Figs.~\ref{fig:setup} and \ref{fig:possibilities})
\begin{align}
 \text{pump pulse:}\quad\quad &W_2(\tau)=0.2\times W(\tau)\,, \label{eq:losses}\\
 \text{probe pulse:}\quad\quad &W_1(\tau,\vartheta_{\rm coll})=r\times\mathfrak{t}_{\tau\to\tau'}(\tau,\vartheta_{\rm coll})\times\mathfrak{t}_{\Delta\omega}(\tau)\times0.5\times0.98^6\times0.2^3\times W(\tau)\,, \nonumber
\end{align}
for the scenario envisioning beam waists of $w_{0,1}^{\rm HM}= w_{0,2}^{\rm HM}=w_0^{\rm HM}=50\,{\rm nm}$. Obviously, for the Setups (a) and (b) without a Molybdenum mirror we have $r=1$.
Also note that the signal derived with \Eqref{eq:dNperpres_approx} eventually has to be multiplied with an overall factor of $0.2\times(0.98)^4$ to obtain the signal registered by the detector in Fig.~\ref{fig:setup}.
Of course, the probe photons traversing the interaction region without vacuum-fluctuation-induced polarization flip, but being scattered into a perpendicularly polarized mode due to the finite purity of polarization filtering are also reduced by this factor.
In line with the above discussion, their total number constituting the background reaching the detector is given by
\begin{equation}
  N_{\rm bgr}={\cal P}\times W_1(\tau,\vartheta_{\rm coll})/\omega\times 0.2\times(0.98)^4\,. \label{eq:Nbgr}
 \end{equation}

For the alternative scenario with beam waists $w_0^{\rm HM}=5\,{\rm nm}$, one has to replace all factors of $0.2\to0.02$ in Eqs.~\eqref{eq:losses} and \eqref{eq:Nbgr}, as well as in the overall factor mentioned in between. The reason for this is the substantial increase in loss ($98\%$ as compared to $80\%$) at the lenses when focusing down to a waist of $w_0^{\rm HM}=5\,{\rm nm}$ instead of $w_0^{\rm HM} =50\,{\rm nm}$ \cite{Huang:2013}.
Finally, we note that the above considerations imply that the use of $\ell$ additional reflections at diamond to fold the beam path to fit the Setups (a)-(c) into a given spatial area would reduce the number of photons available for probing only by a relatively insignificant factor of $0.98^\ell$.

In Tab.~\ref{tab:results_realistic} we provide the explicit values for the attainable numbers of polarization-flipped signal photons accounting for the XFEL pulse deformations and losses detailed here. Both XFEL pulses are assumed to be focused to the same beam waist, i.e., $w_{0,1}=w_{0,2}=w_0$.
This table is the analogue of Tab.~\ref{tab:results_idealized}, which is based on the idealized assumptions of Sec.~\ref{subsec:Idealized} and ignores beam deformations and losses in the experimental setup.
\begin{table}[htbp]
	\centering
	\begin{tabular}{|c|c||*{6}{c|}}
		\hline
		\multirow{2}{*}{Option \#} & \multirow{2}{*}{$w_0^{\rm HM}\,[{\rm nm}]$}  & \multicolumn{2}{c|}{(a)} & \multicolumn{2}{c|}{(b)} & \multicolumn{2}{c|}{(c): $\vartheta_{\rm coll}=179.7^\circ$}\\
		\cline{3-4} \cline{5-6} \cline{7-8}
        & &\multicolumn{1}{c|}{$N_\perp$/h} & \multicolumn{1}{c|}{$N_{\perp,{\rm dis}}$/h} & \multicolumn{1}{c|}{$N_\perp$/h} & \multicolumn{1}{c|}{$N_{\perp,{\rm dis}}$/h} & \multicolumn{1}{c|}{$N_\perp$/h} & \multicolumn{1}{c|}{$N_{\perp,{\rm dis}}$/h} \\
		\hline
		\hline
        1 & \multirow{3}{*}{50} & $1.3\times10^{-9}$ & $9.7\times10^{-17}$ & $4.3\times10^{-8}$ & $4.0 \times 10^{-14}$ & $5.0\times10^{-7}$ & $4.4\times10^{-12}$ \\
		2 & & $5.9\times10^{-6}$ & $5.2\times 10^{-13}$ & $1.5\times10^{-4}$ & $3.4 \times 10^{-10}$ & $1.2\times10^{-2}$ & $1.0\times10^{-6}$ \\
		3 & & $2.4\times10^{-5}$ & $5.7 \times 10^{-12}$ & $1.6\times10^{-3}$ & $3.2 \times 10^{-9}$ & $0.86$ & $6.8\times10^{-5}$ \\
		\hline
		1 & \multirow{3}{*}{5} & $1.3\times10^{-13}$ & $1.2 \times 10^{-19}$ & $9.7\times10^{-12}$ & $6.8\times10^{-18}$ & $1.5\times10^{-9}$ & $2.0\times10^{-14}$ \\
		2 & & $2.1\times10^{-10}$ & $5.2 \times 10^{-17}$ & $1.5\times10^{-8}$ & $3.5\times10^{-14}$ & $1.4\times10^{-5}$ & $0$ \\
		3 & & $2.4\times10^{-9}$ & $5.7 \times 10^{-16}$ & $1.6\times10^{-7}$ & $3.3\times10^{-13}$ & $1.8\times10^{-4}$ & $0$ \\
		\hline
		\multirow{2}{*}
		{4} & 50 & $4.6\times10^{-3}$ & $2.2\times10^{-8}$ & $0.33$ & $1.4\times10^{-5}$  & $14.9$ & $0.019$\\
		& 5  & $4.6\times10^{-7}$ & $2.2\times10^{-12}$ & $3.4\times10^{-5}$ & $1.5\times10^{-9}$ & $0.026$ & $0$\\
		\hline
	\end{tabular}
	\caption{Results for the numbers of signal photons per hour based on the assumptions outlined in detail in Sec.~\ref{subsec:Realistic}.
	For each Setup (a)-(c) we provide results for the different Options 1-4 characterized by different XFEL pulse durations and energies.}
	\label{tab:results_realistic}
\end{table}

All values for the signal photon numbers in Tab.~\ref{tab:results_realistic} are obviously much smaller than the analogous ones in Tab.~\ref{tab:results_idealized}.
Interestingly, for each parameter set the substantial losses coming with the focusing of the probe beam down to $w_0^{\rm HM}=5\,{\rm nm}$ render the signals attainable with the less tight focusing option of $w_0^{\rm HM}=50\,{\rm nm}$ larger than those for $w_0^{\rm HM}=5\,{\rm nm}$.
This even reverses the behavior inferred from Tab.~\ref{tab:results_idealized}.

Out of the Options 1-3 currently available at the European XFEL, now Option 3 with $w_0^{\rm HM}=50\,{\rm nm}$ provides the largest signal photon numbers.
For this case, the total number of probe photons scattered into a perpendicular mode due to the finite purity of polarization filtering is $N_{\rm bgr}\simeq0.06$ photons per shot, which amounts to $\simeq5.6\times10^6$ background photons per hour. This number should be compared with the total number of polarization-flipped signal photons $N_\perp\simeq0.86$.
In line with this, the associated number of discernible signal photons is as small as $N_{\perp,{\rm dis}}\simeq6.8\times10^{-5}$ per hour.

On the other hand, as before Setup (c) which comes closest to the realization of a counter-propagation geometry of the driving laser pulses yields the best results.
Moreover, the projected future Option 4 maximizes the attainable signals.

Interestingly, for Setup (c) the discernible signal now even vanishes for the Options 2, 3, and 4 with $w_0^{\rm HM}=5\,{\rm nm}$. While all other values in Tab.~\ref{tab:results_realistic} are in good agreement with the approximate results predicted by Eqs.~\eqref{eq:dNperpdphiapprox} and \eqref{eq:dNperpdisdphiapprox}, the approximation fails to predict the vanishing discernible signals for the following reason:
In these exceptional cases, the peak value of the signal photon distribution is much smaller than that of the background photons, i.e., ${\rm d}^2N_\perp/({\rm d}\varphi\,{\rm d}\!\cos\vartheta)|_{\vartheta=0}\ll{\cal P}{\rm d}^2N/({\rm d}\varphi\,{\rm d}\!\cos\vartheta)|_{\vartheta=0}$.
In turn, the discernibility criterion~\eqref{eq:discern} could potentially only be met for fairly large values of $\vartheta\ll1$. However, in order to arrive at Eqs.~\eqref{eq:dNperpdphiapprox} and \eqref{eq:dNperpdisdphiapprox}, we have only accounted for the leading term $\sim\vartheta^2$. While a truncation at this order results in an angular regime where the discernibility criterion is superficially met, it turns out that particularly for these cases higher order terms in $\vartheta$ significantly affect the curvature and decay properties of the signal. These modifications prevent the discernibility criterion from being fulfilled for any value of $\vartheta$.

\subsection{Signal Enhancement Strategies}
\label{sec:outlook}

Finally, we highlight several potential enhancement strategies of the signal in our setup that can be within reach in view of contemporary developments of experimental techniques.
\begin{itemize}
    \item[E1] One interesting possibility is the use of two independent XFEL pulses as pump and probe. If, e.g., the pair of CRLs intended to focus and subsequently defocus the XFEL beam to constitute the pump field in Fig.~\ref{fig:setup} could be bypassed by each second XFEL pulse, the signal could be improved: using the XFEL pulses that traverse this pair of CRLs as pump, and those that bypass this pair as probe, the number of pump-probe pulse collisions is reduced by a factor of $1/2$, but at the same time in each collision the signal is increased by a factor of $1/(0.2)^2=25$ for scenarios assuming a pump waist of $w_0^{\rm HM}=50\,{\rm nm}$, and a factor of $1/(0.02)^2=2500$ for scenarios with $w_0^{\rm HM}=5\,{\rm nm}$. In turn, the corresponding results for the attainable signal photon numbers adopting a waist spot size of $w_0^{\rm HM}=50\,{\rm nm}$ ($5\,{\rm nm}$)
follow from those listed in Tab.~\ref{tab:results_realistic} upon multiplication with a factor of $12.5$ ($1250$).
\item[E2] Another means of enhancing the discernible signal to some extend is the use of different focus sizes for the pump and probe, as well as asymmetric focus profiles \cite{Karbstein:2015xra,Karbstein:2016lby}: 
resorting to the refined modeling detailed above we can also consider the scenario where the probe beam ($b=1$) is focused to $w_0^{\rm HM}=50\,{\rm nm}$ and the pump ($b=2$) to $w_0^{\rm HM}=5\,{\rm nm}$. In this case, the loss of each of the two CRLs focusing and defocusing the probe (pump) is about $80\%$ ($98\%$). To account for this in the calculation, the factors of $0.2$ in \Eqref{eq:losses} have to be changed as $0.2\to0.02$ and $0.2^3\to 0.02\times 0.2^2$, respectively. The results for the most prospective Setup (c) with $\vartheta_{\rm coll}=179.7^\circ$ are summarized in Tab.~\ref{tab:asym}.
These findings imply that particularly for Option 3 the asymmetric focusing variant allows to increase the discernible number of signal photons by a factor of $\approx20$ relatively to that attainable by the symmetric focusing variant with $w_0^{\rm HM}=50\,{\rm nm}$ considered in Tab.~\ref{tab:results_realistic}.
\item[E3] Moreover, especially given the recent progress in improving the world record for polarization purity measurements with silicon channel-cut polarimeters to an unprecedented level of accuracy, it is very likely that the polarization purity achievable with diamond quasi-channel-cut polarimeters can be improved by about two orders of magnitude in the near future \cite{Schulze:2018}. This would also come with an enhancement of the discernible signal in our setup. The analytical estimate~\eqref{eq:dNperpdisdphiapprox} allows to infer that for given other parameters and $\vartheta_{\rm coll}\to\pi$ the discernible signal scales with the polarization purity as $N_{\perp,{\rm dis}}\sim{\cal P}^{-(w_{0,2}/w_{0,1})^2/2}$. Hence, in this limit an improvement of $\cal P$ by two orders of magnitude results in an enhancement of the discernible signal by a factor of $10^{(w_{0,2}/w_{0,1})^2}$. While this enhancement factor is as large as $10$ for $w_{0,1}=w_{0,2}$, for the asymmetric focusing option discussed in E2 it becomes just $10^{(1/10)^2}\approx1.02$ and thus essentially does not result in a further enhancement of the signal; cf. also Ref.~\cite{Mosman:2021vua}.
\item[E4] Besides, clearly any means allowing to mitigate the losses of the x-ray lenses used for focusing and defocusing the XFEL beam, such as future technological progress in manufacturing low-loss tailored lens materials, would result in an improved signal. Presently, only speculations about the quantitative gains attainable by such advancements are possible.
\item[E5] Similarly, also an increase of the repetition rate of the XFEL on the $1\,{\rm MHz}$ level or even beyond would be highly beneficial for enhancing the signal. The gain factor resulting from an improvement of the repetition rate from $27000\,\text{pulses/s}$ to $1\,{\rm MHz}$ would be $\approx37$.
\end{itemize}

\begin{table}[h]
\begin{tabular}{|c||c|c|}
\hline
 Option \# & $N_\perp/{\rm h}$ & $N_{\perp,{\rm dis}}/{\rm h}$ \\
 \hline
 \hline
 1 & $7.2\times10^{-9}$ & $6.2\times10^{-9}$\\
 2 & $8.7\times10^{-5}$ & $7.1\times10^{-5}$\\
 3 & $1.9\times10^{-3}$ & $1.3\times10^{-3}$ \\
\hline
 4 & $0.15$ & $0.13$ \\
\hline
\end{tabular}
\caption{Results for the numbers of signal photons per hour based on the refined modeling outlined in detail in Sec.~\ref{subsec:Realistic}. Here, we provide results for Setup (c) with $\vartheta_{\rm coll}=179.7^\circ$. The probe beam is assumed to be focused to $w_0^{\rm HM}=50\,{\rm nm}$, and the pump to $w_0^{\rm HM}=5\,{\rm nm}$.}
\label{tab:asym}
\end{table}

In a last step we assess the prospective size of the discernible signal attainable for idealized XFEL parameters when adopting the more straightforward enhancement strategies E1-E3.
To this end we exclusively focus on Setup (c) with $\vartheta=179.7^\circ\approx\pi$ and the theoretically projected parameters based on state-of-the-art technology~\cite{Chubar:2015mpa} constituting Option 4.
Making use of E1 and E3, we obtain up to
\begin{equation}
    N_{\perp,{\rm dis}}\simeq 0.019 \times \frac{25}{2} \times 10 \simeq 2.37
\end{equation}
discernible signal photons per hour, while employing E1-E3 we arrive at a comparable, but somewhat smaller number of
\begin{equation}
    N_{\perp,{\rm dis}}\simeq 0.13 \times \frac{25}{2} \times 1.02 \simeq 1.66
\end{equation}
discernible signal photons per hour; cf Tabs.~\ref{tab:results_realistic} and \ref{tab:asym}. By means of E5 both of these numbers could, of course, be further increased by an overall factor of $\gtrsim 37$.
Finally, we note that a realistically conceivable integration time of one week of continuous operation would result in an additional enhancement factor of $168$.

\section{Conclusions and Outlook}
\label{sec:concls}

In the present article, we studied the perspectives of an XFEL-only experiment for the first detection of QED vacuum birefringence based on state-of-the-art technology. To this end we utilized a polarization purity of ${\cal P}=1.4\times10^{-10}$ \cite{Bernhardt:2020vxa}.
Our study is the first to consistently account for the losses of the focusing optics as well as pulse deformations induced by reflections of the original XFEL pulse at crystal surfaces.
In particular, we have demonstrated how severely predictions based on seemingly reasonably realistic assumptions may be affected when accounting for the details of the experimental setup devised to actually detect the phenomenon. 
This experiment has certainly many challenges: the small submicron focus size requires very stable conditions of both the focusing and the beam optics.
Moreover, the pointing of the XFEL needs to be extremely high and has to be controlled for each pulse.
Also the requirements for the spatio-temporal overlap of the focused pump and probe beams are enormous and should be controlled for each laser pulse train.
Specifically the fluctuation of the temporal delay must be significantly shorter than the pulse length.
In addition, the employed diamond crystals must provide sufficient crystal perfection in order to deliver high Bragg crystal reflectivity.
At the same time, the x-ray focusing optics should deliver highly perfect focus properties close to the diffraction limit and have no influence of the highly polarized state of the beam.
All these properties need very stable temperature condition as well as low vibrational movements of all optics.

The attainable signal depends on the brightness of the source. 
In the experimental scenario envisioned by us the brightness of the source is clearly an essential key parameter: on the one hand, it determines the intensity of the pump field driving the vacuum birefringence effect. On the other hand, it controls the number of photons available for probing the phenomenon.
To this end, we note that prospective XFEL-oscillators which are currently under discussion~\cite{Adams:2019sci} will surpass the brightness of current XFELs by orders of magnitudes.
Our study clearly identifies those building blocks that limit the quality of an experiment most severely -- most prominently the sizable losses at the x-ray lenses. Improvements of these components could significantly increase the feasibility of such experiments.
Moreover, we emphasize that future high-precision x-ray polarimeters have the potential to reach polarization purities ${\cal P}\lesssim10^{-12}$~\cite{Schulze:2018}, which would generically also increase the discernible signal.

Such developments will substantially improve the perspectives for precision tests of QED vacuum birefringence in XFEL based laboratory experiments.

\acknowledgments

This work has been funded by the Deutsche Forschungsgemeinschaft (DFG) under Grant Nos. 416607684; 416611371; 416700351 within the Research Unit FOR2783/1.

\end{document}